\documentclass[a4paper,fleqn,usenatbib]{mn2e}
\usepackage{newtxtext,newtxmath}
\usepackage[T1]{fontenc}
\usepackage{ae,aecompl}
\usepackage{amssymb,amsmath}
\usepackage{natbib}
\usepackage{multirow}
\usepackage[english]{babel}
\usepackage{graphicx}
\usepackage{color}
\usepackage{sidecap}
\usepackage{comment}
\usepackage{epstopdf}
\definecolor{MyGreen}{rgb}{0.0,0.5,0.4}
\definecolor{MyRed}{rgb}{1.0,0.4,0.0}
\definecolor{Gray}{rgb}{0.6,0.6,0.6}

\newcommand{\amir}{\textcolor{black}}
\newcommand{\amiradd}{\textcolor{black}}
\newcommand{\amiraddF}{\textcolor{black}}

\newcommand{\pd}{\partial}
\newcommand{\fnull}{\text{null}}
\newcommand{\fexp}{\text{exp}}
\newcommand{\fdiff}{\text{diff}}
\newcommand{\os}{\text{$1$step}}
\newcommand{\Tsim}{t_{\text{sim}}}
\newcommand{\Jsim}{j_{\text{sim}}}
\newcommand{\lar}{\text{large}}
\newcommand{\delj}{j_\delta}
\newcommand{\deljv}{\vec{\delj}}
\newcommand{\sma}{\text{small}}
\newcommand{\jscat}{j_{\text{scat}}}
\newcommand{\jtarg}{j_{\text{targ}}}
\newcommand{\LC}{\text{lc}}
\newcommand{\BH}{\text{bh}}
\newcommand{\Lambzero}{\Lambda_{\text{lc}}}
\newcommand{\Lambmax}{\delta\Lambda^{\max}_{\text{lc}}}
\newcommand{\circular}{\text{cir}}
\newcommand{\normcir}{\text{cir}}

\fontsize{13pt}{10pt}

\title[How empty is an empty loss cone?]{How empty is an empty loss cone?}

\author[A. Weissbein and R. Sari]{
	Amir Weissbein,$^{1}$\thanks{E-mail: amir.weissbein@mail.huji.ac.il}
	Re'em Sari,$^{1}$
	\\
	$^{1}$Racah Institute of Physics, Hebrew University, Jerusalem 91904, Israel\\
}

\date{Accepted 22/2/2017. Received 21/2/2017; in original form 25/8/2016}

\pubyear{2016}

\begin{document}
	
\label{firstpage}
\pagerange{\pageref{firstpage}--\pageref{lastpage}}
\maketitle

\begin{abstract}
We consider two body relaxation in a spherical system with a loss cone. Considering two-dimensional angular momentum space, we focus on "empty loss cone" systems, where the typical scattering during a dynamical time $j_{d}$ is smaller than the size of the loss cone $j_{\LC}$. 
As a result, the occupation number within the loss cone is significantly smaller than outside. Classical diffusive treatment of this regime predict exponentially small occupation number deep in the loss cone.

We revisit this classical derivation of occupancy distribution of objects in the empty loss cone regime. We emphasize the role of the rare large scatterings and show that the occupancy does not decay exponentially within the loss cone, but it is rather flat, with a typical value $\sim [(j_d/j_{\LC})]^2\ln^{-2}(j_{\LC}/j_{\min})$ compared to the occupation in circular angular momentum (where $j_{\min}$ is the smallest possible scattering). 

Implication are that although the loss cone for tidal break of Giants or binaries is typically empty, tidal events which occurs significantly inside the loss cone ($\beta\gtrsim 2$), are almost as common as those with $\beta\cong 1$ where $\beta$ is the ratio between the tidal radius and the periastron. The probability for event with penetration factor $>\beta$ decreases only as $\beta^{-1}$ rather than exponentially. This effect has no influence on events characterized by full loss cone, such as tidal disruption event of $\sim 1m_\odot$ main sequence star. 
\end{abstract}


\section{Introduction}

Many physical phenomena, especially in the dynamics around the galactic center involve destructive processes.
Perhaps the simplest problem is that of tidal disruption of a star when it is close enough to the galactic center massive black hole.
Orbits which penetrate that distance are called loss cone orbits and are characterized by low specific angular momentum.
If orbital elements of objects around the galactic center black hole were fixed, all loss cone orbits have already been destroyed and the loss cone would be completely empty.
   
Many previous works have discussed the loss cone filling rate 
\citep[e.g.][]{FR1976,LS1977,CK1978,MT1999,Yu2002,YT2003,MM2003}, and see \cite{MerrittBook2013}
for a comprehensive discussion.
Most of these works assume that loss cone filling
is well described by the Fokker-Planck equation, i.e by diffusion in the angular momentum - energy phase space.
The rational is that if the typical change in energy or angular momentum during a dynamical time are small compared to the size
of the loss cone, diffusion is an accurate description.
However, it is well know that hard scatterings contributes equally to the diffusion coefficient. Such scatterings are not captured correctly
by diffusion or Fokker-Plank equation \citep{BKA2013}. 
	
To address that, we start in section \S 2 with a Boltzmann equation, which correctly describes small as well as large scatterings.
We go over a series of possible approximate solution in \S 3. Finally in \S 4 we compare those approximation with numerical
experiments which simulate the full Boltzmann equation.
 
\section{The Boltzmann equation with loss cone}\label{BoltzmannEquation}
We consider the evolution due to two body scattering of the otherwise conserved values of the stellar orbits, i.e the angular 
momentum $\vec{j}$ and the energy $E$.
The loss cone is represented by $||\vec{j}||<j_{\LC}(E)$, where $j_{\LC}(E)$ is the angular momentum related to some critical periastron. 
  
Since scatterings driven by gravitational interactions makes stars enter the loss cone mainly due to angular momentum loss \citep{LS1977,MerrittBook2013},
we treat the whole dynamics as two dimensional $\vec{j}$ phase space 
(two tangential components of the angular momentum), and ignore the energy. 
Moreover, since the system have polar symmetry, the distribution function only depends on the magnitude of $\vec{j}$
($0\leq j \lesssim j_{\circular}$).

We define the occupancy distribution $f(t,\vec{j})$
to be proportional to the number of stars at time $t$, 
per unit angular momentum square around angular momentum $\vec{j}$,
i.e $f(t,\vec{j})d^2\vec{j}$ is proportional to the number of stars with angular momentum $\vec{j}$. Therefore $f$ has units of $1/j^2$.
We use the normalization $f(j_{\circular})\equiv f_{\normcir}$
where $f_{\normcir}$ is approximately the total number of objects in the system over 
		$4\pi j_\circular^2 $.

We account for two processes: scattering and destruction, and focus on orbits with low angular momentum compared to that of a circular orbit. This is an excellent assumption for the galactic center where loss cone orbits have angular momentum over a thousand time smaller than those of circular orbits.
We assume that the rate of scattering by an amount $R(\deljv)$, is independent of the current angular momentum of the object .
This is justified since for the almost radial orbits most of the scattering occurs far from pericenter and the exact position of the pericenter is thus unimportant.
The destruction process, eliminates once every dynamical time objects of low enough angular momentum $j<j_{\LC}$. 
Exact description of such process requires following the orbital phase of each star. To avoid that, and have the distribution function 
depend on angular momentum only, we use the following approximate Boltzmann equation:
\begin{equation}\label{GeneralBoltz}
\begin{array}{l}
\frac{\pd f(\vec{j})}{\pd t}
=-(t_d/2)^{-1}f(\vec{j})\Theta(j_{\LC}-j) +\\
\\
\int d^2\deljv\  \left[f(\vec{j}-\deljv)-f(\vec{j})\right]R(\deljv) .
\end{array}
\end{equation}
where $t_d$ is the orbital time, or dynamical time, of the particles,
$\Theta(j_{\LC}-j)$ is the Heaviside function which gets $1$ for $j<j_{\LC}$ and $0$ for  $j>j_{\LC}$
and $R(\deljv)$ is the differential rate of scatterings of size and direction $\deljv$,
i.e. number of scattering per unit time per $d^2\deljv$.  

The sink term $-(t_d/2)^{-1}f(\vec{j})\Theta(j_{\LC}-j)$, which takes the loss process into account, and especially the factor of $1/2$, warrants some discussion. 
A star may get scattered into the loss cone, $j<j_{\LC}$, at any phase of its orbit. 
It will be lost when its orbital phase brings it close enough to the central object. 
To mimic that without following the phases of individual stars, 
we have to make an assumption on the phase distribution for particles in the loss cone.

If scattering in and out of the loss cone is faster than the orbital time, the loss cone will be uniformly populated in orbital phase. In such a case,
the rate of destruction of particles is equal to the number of particles in the loss cone divided by the orbital time. Therefore, the sink term
in this limit is $-t_d^{-1}f(\vec{j})\Theta(j_{\LC}-j) $.

However, we are interested in the opposite limit, where scattering deep into the loss cone
are rare.
In this ``empy loss cone" limit, a particle that was scattered deep into the loss cone will remain there until it gets destroyed 
If we assume that it entered the loss cone with
an arbitrary phase, then a star survives within the loss cone for half its orbital period on average. The sink term is therefore given by
$-(t_d/2)^{-1}f(\vec{j})\Theta(j_{\LC}-j) $, which is what we used in our equation \ref{GeneralBoltz}.

\amir{
	Our choice of the sink term,
	mimics well the behavior deep inside the LC and  
	allows us to find the occupancy distribution deep inside the LC.    
	Close to the LC's edges on the other hand, 
	where the typical star may scattered inward and outward of the LC during one dynamical time, 
	this sink term differs from the real physical behavior by numerical factor of order $\sim 2$. 
	We therefore expects the calculated occupancy distribution to be a good approximation inside the LC, 
	and to err by some on its edges.     
	}

\amir{We find the scattering's rate term $R(\deljv)$.}
We consider small LC, i.e. $j_{\LC}$ much smaller than the circular angular momentum  $j_{\circular}$ .
This is the relevant case for HVSs or tidal disruption events.
In this case, the stars which are about to get scattered into the loss cone,
are moving almost radially with respect to the central BH during most of their period. 
Their scatterers on the other hand, are the more numerous stars moving on much more circular orbits.
%
%
%
%
%
\amir{
	In this case, of gravitational scatterings of stars with low angular momentums, 
	the velocity's change is in the direction perpendicular to the BH's direction.
	Therefore $\delj\equiv a\Delta v_\perp\simeq Gma/bv $. 
	where here 
	$\Delta v_\perp$ is the velocity change in the perpendicular direction,
	$m$ is the mass of an individual scatterer,
	$a$ is the distance from the central BH and 
	$v$ the relative velocity of the encountering objects.	
	The rate of such encounters is proportional to $b^2$, so that the rate of encounters resulting in
	angular momentum change of order $\delj$ is proportional to $\delj^{-2}$. 
}

\amir{
	Moreover, since stars only interact with stars in their close surrounding, 
	we may assume that in any phase of the orbit, the subject star interact with an isotropic background. 
	As a result, the probability that an individual subject star experience a scattering due to an impact parameter $b$, does not depend on the direction.
	Thus,
	$R(\deljv)$ which is the rate per unit $d^2\deljv$ is isotropic and proportional to 
	$||\deljv||^{-4}$. 
	}
We therefore define 
\begin{equation}\label{Rj}
R(j)=R_0 |\deljv|^{-4}\ \text{ for }\ j_{\min}<|\deljv|< j_{\max} \quad , 
\end{equation}
where $R_0$ is some function of the masses of the particles and the velocity dispersion
and $j_{\min},\ j_{\max}$ are the minimal and the maximal possible scattering: $j_{\min}$ corresponds to an impact parameters of order the distance between stars,
and $j_{\max} \approx j_\circular$ corresponds to an impact parameter that results in a velocity change of order unity

Substitute the rate term (equation \ref{Rj}) into the Boltzmann equation (equation \ref{GeneralBoltz}) our
Boltzmann equation becomes 

\begin{equation}\label{Boltz}
\begin{array}{l}
\frac{\pd f(\vec{j})}{\pd t}=
-2t_d^{-1} f(\vec{j})\Theta(j_{\LC}-j) +\\ \\
R_0 \int d^2\deljv\  |\deljv|^{-4}\left[f(\vec{j}-\deljv)-f(\vec{j})\right]
\end{array}
\end{equation}

It is the process described by this equation that we simulate numerically in \S 4.

\section{Solution to the Boltzmann equation}
The system we describe has no steady state solution, since it contains a sink (the loss cone) without any specified supply term.
We may force a steady state solution to the Boltzmann equation by normalizing
		$f(j_\circular)\equiv f_{\normcir}$ independent of time.
Physically this could be thought as a result of supply of new objects to the outskirts of the loss cone, e.g. from orbits of higher energy which we do not consider, or an approximation of low angular momenta $j \ll j_{\circular}$ whose intrinsic evolution should have been fast.

In this section, we list approximate solutions to equation \ref{Boltz} with different levels of accuracy.
We denote the solution sunder these approximate solutions by 
$f_{\fnull}(j)$, $f_{\fexp}(j)$, $f_{\fdiff}(j)$ and $f_{\os}(j)$
while the colors: blue (dashed), green (dashed), orange and red are used to mark 
those four solutions respectively in all figures.  

\subsection{Diffusion with null loss cone - $f_{\fnull}(j)$ }\label{chapfdiff}
		
The diffusion approximation assumes that scatterings can be considered small, i.e., 
the distribution function $f(\vec{j})$ does not change significantly over the size of the scattering. 
In this case the integrand on the right hand side of equation \ref{Boltz} 
may be expanded using Taylor approximation and equation \ref{Boltz} 
for $j>j_{\LC}$ may be rewritten as
\begin{equation}\label{Basic_Null}
\frac{\pd f}{\pd t}= 
\frac{1}{j}\frac{\pd}{\pd j}\left(j\cdot D_{\max}\frac{\pd f_{\fnull}}{\pd j}\right)
\end{equation}
where:
\begin{equation}
D_{\max}\equiv \frac{1}{4}\int d^2\deljv\  |\delj|^2 R(\delj)={\pi \over 2} R_0\Lambda_{\max} 
\end{equation}
Where $\Lambda_{\max}\equiv\ln(j_{\max}/j_{\min})$
\footnote{\label{LambdaDef}
	We use the notaion $\Lambda_i$ to define 
	$\ln(j_i/j_{\min})$ at other cases as well: 
	$\Lambda_{d}\equiv \ln(j_{d}/j_{\min})$, 
	$\Lambda_{\LC}\equiv \ln(j_{\LC}/j_{\min})$,
	$\Lambda_{\circular}=\ln(j_{\circular}/j_{\min})$.	
	and 
	$\Lambda_{\max}=\ln(j_{\max}/j_{\min})$.
	We also use the notation $\Lambmax$ to define $\ln(j_{\max}/j_{\LC})\equiv\Lambda_{\max}-\Lambda_{\LC}$
}.
The expression for the diffusion coefficient $D_{\max}$ shows that scattering of all scale contribute equally to the diffusion, 
leading to the familiar coulomb logarithm $\Lambda_{\max}=\ln(j_{\max}/j_{\min})$. 

We find the steady state solution of equation \ref{Basic_Null}.
In systems where the typical scattering over the dynamical time is smaller than the size of the loss cone, 
the loss cone is relatively empty. 
In the current approximation we therefore take the boundary condition
$f_{\fnull}(j_{0})=0$, where  $j_{\LC}-j_0\ll j_{\LC}$.
\citep{LS1977,MT1999,MerrittBook2013}. 

This approximation gives us a steady state solution of the form
\begin{equation}\label{fDiffusion}
f_{\fnull}(j)=\left\{
\begin{array}{c c}
0 & j<j_{0} 
\\
f_{\normcir}(\Lambmax)^{-1}\ln(j/j_{0})  & j>j_{0} 
\end{array}
\right.\ .
\end{equation} 
where 
\begin{equation}
\begin{array}{l l}
\Lambmax\equiv&\Lambda_{\max}-\Lambda_{\LC}=\\
& \ln(j_{\max}/j_{0})\simeq \ln(j_{\max}/j_{\LC})
\end{array}
\end{equation}
The solution of the form $f(j)\propto \ln(j/j_0)$
was found by \cite{LS1977} (see also \cite{MerrittBook2013}).
The blue dashed line in figures \ref{figsimA},\ref{figsimB} and \ref{figsimC} marks $f_{\fnull}(j)$. 
Here, we normalized $f$ to $f_{\normcir}$    
	at some large angular momentum $j_{\max}$.
	Typically $j_{\max}\sim j_{\circular}$, if more than the relaxation time has elapsed and $j_{\max}\sim (Dt)^{1/2}$ at earlier times.

This solution describes a constant flux of objects towards the loss cone, given by 
$\pi^2R_0f_{\normcir}\Lambda_{\max}/\Lambmax$. 
However, it was derived using the zero boundary conditions.
By using this boundary condition, we give up the ability to inspect the distribution of particles within the loss cone. 
The question of how empty is an empty loss cone,
how deep particles penetrate into it (what is exactly $j_{\LC}-j_0$?)
and what is the distribution of $j$ of particles that suffer distruction can not be answered in this framework.

\subsection{Diffusion with Delayed Distractio - $f_{\fexp}$ \label{diffdelay}}

The first correction, which was already used by \cite{CK1978},
allows to account for particles inside the loss cone.
We drop the null boundary condition $f(j_{\LC})=0$, and instead take a sink term inside the loss cone just as in the original Boltzmann equation (eq. \ref{Boltz}). The resulting diffusion equation for steady state is
\begin{equation}\label{temp}
\begin{array}{c}
0=
\frac{1}{j}\frac{\pd}{\pd j}\left(j\cdot D_{\max}\frac{\pd f_{\fexp}}{\pd j}\right) -
2t_d^{-1}f_{\fexp}(j)\Theta(j_{\LC}-j)\\
\end{array}
\end{equation}
This could be easily solved both inside and outside the loss cone.
The two solutions are then matched to ensure continuity of
$f$ and $\pd f/\pd j$ at $j_{\LC}$.
The solution inside the loss cone is given by a Bessel function,
but since it decays so strongly within the loss cone,
we ignore the cylindrical geometry and approximate the Bessel function as an exponential.
We obtain
\begin{equation}\label{fdelay}
f_{\fexp}(j)=\left\{
\begin{array}{c c}
f_{\normcir}
\cdot\frac
{\frac{j_2}{j_{\LC}}\exp\left\{\frac{j-j_{\LC}}{j_2}\right\}}
{\left[\frac{j_2}{j_{\LC}}+\Lambmax\right]}
& j<j_{\LC} 
\\
\\
f_{\normcir}
\cdot\frac
{\left[\frac{j_2}{j_{\LC}}+\ln\left(\frac{j}{j_{\LC}}\right)\right]}
{\left[\frac{j_2}{j_{\LC}}+\Lambmax\right]}
& j>j_{\LC} 
\end{array}
\right.
\end{equation} 
with
\begin{equation}
j_{2}\equiv (t_dD_{\max}/2)^{1/2} \ \ .
\end{equation}
Note that the parameter $j_2$ is alike \cite{LS1977}'s $j_2$ which they define as 
$(t_d/t_{\text{relax}})^{1/2}j_{\circular}$, since $t_{\text{relax}}=j_{\circular}^2/D_{\max}$.
	Moreover,
	$f_{\fexp}(j)$ outside the LC, 
	may be written as $\ln(j/j_0)$, 
	where $j_0\equiv j_{\LC}\exp(-j_2/j_{\LC})\cong j_{\LC}-j_2$, 
	i.e $j_2$ is the typical penetration of particles into the LC.     
The green dashed line in figures \ref{figsimA},\ref{figsimB} and \ref{figsimC} marks $f_{\fexp}(j)$. 

We compare our occupancy $f_{\fexp}$  close to the loss cone's edge with those of \cite{CK1978}
in the empty LC regime $j_2\ll j_{\LC}$.
The functional form of the solution is similar.
However, their penetration of particles into the loss cone 
($j_{\LC}-j_0$ where $j_0$ is set according to the $\propto\ln(j/j_0)$ behavior of the profile outside the LC)
is larger by a numerical factor of $1.165$
than ours\footnote{Milosavljevic solution of \cite{CK1978}'s  Fokker-Planck equation as presented in \cite{MerrittBook2013}, results in a factor of $(\pi/2)^{1/2} \cong 1.25$ }.
The reason for the difference is that at depth of $\lesssim j_2$ inside the loss cone, 
our assumption that the average time that a particle spends inside the loss cone is $t_d/2$ (used in equation \ref{Boltz} and \ref{fdelay}, and therefore in the definition of $j_2$ above) is not accurate (see discussion below equation \ref{GeneralBoltz}).
A star that resides very close to the boundary of the loss cone, has a chance to be scattered in and out of it within a dynamical time. Their phases are therefore more uniform and the factor of $2$ we have used should be closer to unity. 

\subsection{$f_{\fdiff}(j)$ - Variable diffusion coeficient}\label{chapfdibar}\label{chapfdelay}

The diffusion approximation is valid as long as the steps size is smaller than the scale over which the function $f$ changes.
Our solution in \S\ref{diffdelay} shows that far outside the loss cone, the scale over which the function $f$ changes is roughly the distance from the loss cone, while very close to the loss cone edge, it changes on 
a scale $j_2$.
\amir{
	Higher orders of the Fokker-Planck equation can partially account for this effect \citep{BKA2013}. However, where these effects are strong, infinite numbers of terms would need to be considered. We therefore take another approach.}
In appendix \ref{appendix_large_scatterings}, we argue that outside the loss cone, large scatterings bigger than this scale, turn out to be less important. 

\amir{Therefore,} taking into account only the relevant small scatterings, 
reduces the coulomb logarithm that appears in
the diffusion coefficient. This leads as to \amir{an effective}
 diffusion equation with a variable diffusion coefficient:
\begin{equation}\label{eqf0}
0= 
\frac{1}{j}\frac{\pd}{\pd j}\left(jD(j)\frac{\pd f_{\fdiff}}{\pd j}\right) - 
2t_d^{-1}f_{\fdiff}(j)\Theta(j_{\LC}-j)
\end{equation}
with
\begin{equation}\label{Dfuncj}
D(j)\equiv \frac{\pi }{2} R_0
 \ln\left(\frac{\max\{j - j_{\LC},j_{d}\} }{j_{\min}}\right)     
\end{equation} 
where $j_d$ is the scale on which $f(j)$ is changing close to the LC surrounding.
Notice that due to the logarithmic dependency of $D(j)$, 
$f(j)$ close to the LC is changing on scale shorter than $j_2= (t_dD_{\max}/2)^{1/2}$
and therefore $j_d<j_2$. 
We define $j_d\equiv (t_dD(2j_{\LC})/2)^{1/2}$ or
\begin{equation}\label{jd_def}
j_d\equiv 
\left[
\frac{\pi}{4} t_d R_0
\ln\left(\frac{j_{\LC}}{j_{\min}}\right)
\right]^{1/2}
\end{equation}	   	

In this description,  
we have not focused on the best estimate for the diffusion coefficient inside the loss cone,
since as we shall see in the next section, the loss cone is mostly populated by non diffusive processes.

Solving this equation inside and outside the loss cone under continuity conditions,
we find the steady state solution to this equation to be approximately
	\begin{equation}\label{fdibar}
	\begin{array}{l}
	f_{\fdiff}(j)=
	\\
	\\
	\left\{
	\begin{array}{c c}
	f_{\normcir}
	\cdot\frac
	{\frac{j_d}{j_{\LC}}\Lambda_{\LC}^{-1}
		\exp\left\{\frac{j-j_{\LC}}{j_d}\right\}}
	{\frac{j_d}{j_{\LC}}\Lambda_{\LC}^{-1}+\ln
	\left[1+\Lambda_{\LC}^{-1}\Lambmax\right]}		
	& j<j_{\LC}
	\\
	\\
	f_{\normcir}\cdot\frac
	{\frac{j_d}{j_{\LC}}\Lambda_{\LC}^{-1}+\ln
		\left[1+\Lambda_{\LC}^{-1}\ln\left(\frac{j}{j_{\LC}}\right)\right]}
	{\frac{j_d}{j_{\LC}}\Lambda_{\LC}^{-1}+\ln
		\left[1+\Lambda_{\LC}^{-1}\Lambmax\right]}		
	& j>j_{\LC}
	\end{array}
	\right.
	\end{array}
	\end{equation} 
Our solution \ref{fdibar} is approximate. It is an exact solution to equation \ref{eqf0} with diffusion coefficient of the form 
$D(j)=(\pi R_0/2)\cdot\ln(j/j_{\min})$, rather than the diffusion coefficient presented in \ref{Dfuncj}.
As a result, the solution is inaccuraate very close to the LC edge 
where $0<|j-j_{\LC}|\lesssim (j_{\LC}j_{\min})^{1/2}$.
It is, however, quite accurate elsewhere.  
The orange line in figures \ref{figsimA},\ref{figsimB} and \ref{figsimC} marks $f_{\fdiff}(j)$. 

Note that very far from the loss cone,
where $\ln(j/j_{\LC})\gg \Lambzero$,
$f_{\fdiff}(j)$ profile satisfy $f(j)\propto\ln[\ln(j/j_{\LC})]$ rather than $f(j)\propto\ln(j/j_{\LC})$.
This flattening of the profile compared to the expectation from constant diffusion coefficient ($f_{\fexp}(j)$),
comes from the logarithmic reduction of the diffusion coefficient at low $j$'s: 
$D\sim R_0\ln(j_{\LC}/j_{\min})$ rather than 
$D\sim R_0\ln(j_{\max}/j_{\min})$. 
This reduction, dictate lower flux also at large $j$'s where 
$D(j)\sim D(j_{\circular})\sim D_{\max}$. 

For similar normalization of $f(j_{\circular})\equiv f_{\normcir}$, 
\amir{we find}
the ratio between 
the flux according to $f_{\fdiff}$ and the flux according to $f_{\fexp}$
\amir{by comparing 
$\pd f_{\fdiff}/\pd j|_{j_{\circular}}$
and
$\pd f_{\fexp}/\pd j|_{j_{\circular}}$.
We obtain: 
} 

\begin{equation}\label{rate_factor}
	\frac
	{\left.\frac{\pd f_{\fdiff}}{\pd j}\right|_{j_\circular}}
	{\left.\frac{\pd f_{\fexp}}{\pd j}\right|_{j_\circular}}
	=
	\frac
	{\Lambmax}
	{\Lambda_{\max}\ln\left[1+\left(\frac{\Lambmax}{\Lambzero}\right)\right]}				
\end{equation}
This reduction can never be more significant than order unity. 
If $\Lambmax\ll \Lambzero$, the reduction is simply the $\Lambzero/\Lambda_{\max}$ 
equivalent to the diffusion coefficient reduction in the LC surroundings.
If on the other hand $\Lambmax\gg \Lambzero$, the reduction is $\sim \ln(\Lambda_{\max}/\Lambzero)$ - which for realistic parameter can not be larger than $\sim 3$.

Specific examples are given in \S\ref{realgalaxy}.
 
\subsection{$f_{\os}$ - The effect of large scatterings}\label{chapf1step}

The delayed destruction by the loss cone, predicts exponentially small amount of particles deep inside the loss cone.
The typical decay length of the exponent, 
is $j_d$, the typical diffusion during half dynamical time (see equation \ref{fdelay}),
which is necessarily $\ll j_{\LC}$ in the case of empty LC. 
On the other hand, scatterings of size $j\sim j_{\LC} \gg j_d$ are rare only by factor $\sim(j_{\LC}/j_d)^2$ compared to $j_d$ scatterings.
Via those scatterings, particle may enter anywhere in the loss cone with roughly constant probability. 
Therefore, we expect that inside the loss cone, at depth larger than $\sim j_d$, the profile is dominated by particles that entered into the loss cone due to one large scattering rather than by diffusion on time scales $\sim t_d$. 

We define $f_{\os}(j)$ as the steady state solution $f_{\fdiff}(j)$ evolved by a {\it single} step taken from the probability distribution of steps over time $t_d/2$ (the average time particles survive inside the loss cone). 
Since the fraction of particles that preform scattering of size and direction $\deljv$ during time $t_d/2$, is $R(\deljv)t_d/2$ (true only for $j\gg j_d$),
we find that deep inside the loss cone, at $j$'s satisfies $j_{\LC}-j\gg j_d$, the profile is: 
\begin{equation}\label{f1step}
\begin{array}{c}
f_{\os}(j)=\frac{t_d}{2}\int_{j_d}^{j_{\max}} d^2\deljv\  R(\deljv)f_{\fdiff}(\vec{j}-\deljv)=
\\
\\
\frac{t_d R_0}{2}\int_{j_d}^{j_{\max}} d^2\deljv\  |\deljv|^{-4}f_{\fdiff}(\vec{j}-\deljv)
\end{array}
\end{equation}   
This equation may be solved numerically everywhere inside the loss cone and this solution noted by the red dashed line in figures \ref{figsimA}, \ref{figsimB} and \ref{figsimC}.

Specifically, in the central part of the loss cone ($j \ll j_{\LC}$),
\amir{one may estimate $f_{\os}(j)$, by substituting $j=0$ in the integrand and solve the integrals. 
Than, in first order in $\Lambda_{\LC}^{-1}$ and $j_d/j_{\LC}\ll 1$ }
the expression for $f_{\os}$ may be approximated as:   
\begin{equation}\label{f1stepdeep}
\begin{array}{l l}
f_{\os}^{deep}&\cong  
f_{\normcir}\left(\frac{j_d}{\Lambda_{\LC}j_{\LC}}\right)^2
\frac{\left[1-\frac{1}{2\Lambzero}+\frac{2j_d}{j_{\LC}}\right]}
{\frac{j_d}{j_{\LC}}\Lambda_{\LC}^{-1}+\ln[1+\Lambzero^{-1}\Lambmax]} 
\\ 
&\sim \left(\frac{j_d}{\Lambda_{\LC}j_{\LC}}\right)^2 f_{\normcir}
\end{array}
\end{equation}
This value is marked by the horizontal dashed-doted black line in figures \ref{figsimA} and \ref{figsimC}.

One major consequence of the last result is that the number of objects in the deep LC 
(say $j<j_{\LC}/2$) is lower only by a factor of order unity than the number of objects on the 
LC's edge: 
\begin{equation}
\frac{\pi \left(\frac{j_{\LC}}{2}\right)^2\cdot f_{\os}^{deep}}{2\pi j_{\LC}j_d \cdot f_{\fdiff}(j_{\LC})}\sim
\frac{1}{8\Lambzero}
\end{equation}
instead of the more significant factor
\begin{equation}
\frac
{\pi \left(\frac{j_{\LC}}{2}\right)^2\cdot f_{\fdiff}(j=1/2)}
{2\pi j_{\LC}j_d \cdot f_{\fdiff}(j_{\LC})}
\sim
\frac{j_{\LC}}{8j_d} \exp\left\{-\frac{j_{\LC}}{2j_d}\right\}
\end{equation}
predicted by diffusive model. 


\subsection{Analytical solutions - Summary}
Although we could not find an exact solution to equation \ref{Boltz}, 
we provide approximate solutions in different regimes.

Outside the loss cone and at depth of $\lesssim j_d\equiv [(\pi/4)t_dR_0\ln(j_{\LC}/j_{\min})]^{1/2}$ 
into the loss cone the profile is dominated by diffusive behavior with delayed destruction.
Therefore the solution is approximately $f_{\fdiff}(j)$ (equation \ref{fdibar}).
Deeper than $\sim j_d$ inside the loss cone, only large scatterings contribute to the population and therefore the solution is approximately $f_{\os}(j)$ (equation \ref{f1step}).

%
Since this solution takes into account large scatterings which are not considered in a Fokker-Planck equation, it is qualitatively different than previous solutions.
In particular:
\begin{itemize}
	\item {
		The profile far outside the LC, in angular momentums $j\gg j_{\LC}^2/j_{\min}$ 
		(if such angular momentums are yet $\lesssim j_{\circular}$), is not $\propto \ln(j/j_{\LC})$, but
		\amiradd{flatter, with asymptotic behavior $\propto \ln[\ln(j/j_{\LC})]$ (equation \ref{fdibar}).}
		As a result, the LC filling rate is reduced by logarithmic factor (equation \ref{rate_factor})
		}
	\item{
		The profile deep inside the LC is not exponentially small with a scale of $j_d$, 
		but rather polynomially $\sim [j_d/j_{\LC}\ln(j_{\LC}/j_{\min})]^2f_{\normcir}$. 
		}
\end{itemize} 

Special attention to large scatterings was already discussed in \cite{BKA2013}. 
\cite{BKA2013}
focussed on energy rather than angular momentum relaxation. 
They used higher order terms of the Master equation (or Boltzmann equation) than those leading to the familiar Fokker-Planck equation.

\section{Numerical Experiments}\label{numerical}
We compare our analytical results to numerical Monte-Carlo simulation results in
order to demonstrate that $f_{\fdiff}(j)$ and $f_{\os}(j)$ are more accurate than the previous solutions.
In this simulation, we let a large number of particles $N_p$ evolve over time from a flat distribution to an effective steady state distribution. 
Each particle starts from some random location in $j$'s space and earn random scatterings over long time.
We run this simulation over large virtual time $\Tsim$, 
so that a quasi steady state is achieved in some finite region $j<\Jsim\equiv [\Tsim D(2j_{\LC})]^{1/2}$ while $j_{\LC}\ll \Jsim$).

\subsection{Simplified simulation description}\label{simple}

We employ the following simple algorithm for finding the steady state solution of equation \ref{Boltz} in low angular momentums ($j<\Jsim$).
In practice, since this algorithm is slow, we employ an accelerated version of it
as described in appendix \S\ref{improve}. We presented here the simplified (non accelerated) algorithm for simplicity.

\begin{itemize}
\item {
A single point particle has initial angular momentum $j<j_{\max}$ and a random direction. 
The probability for the particle to start in a ring $\{j,j+dj\}$ is proportional to $j$.
}
\item {
Every step each particle preform a single jump.
The direction of the jump is randomly picked with equal probability to each direction.
The size of the jump $j$ varies between $j_{\min}$ and $j_{\max}$ with probability $\propto j^{-3}$ (corresponding to $R(\deljv)\propto \delj^{-4}$ - equation \ref{Rj}). We change the position of the particle according to the chosen size and direction of the jump.

After each jump, the time is advanced by a constant
\begin{equation}
t_{\min} \equiv\left[\int_{j_{\min}}^{j_{\max}} d^2\deljv \ R(\delj)\right]^{-1}
\end{equation}
which correspond to the time it takes to a particle to make a step larger than $j_{\min}$ according to equation \ref{Rj}.
}


\item {
Whenever a particle has penetrated to the loss cone, i.e. 
it has $j<j_{lc}$,
it has a probability of $1-e^{-2t_{\min}/t_d}$ to be expelled.
This mimics the loss cone term which appears in the Boltzmann equation (equation \ref{Boltz}).
If the particle has been expelled, it does not enter into the statistics of final locations.
\amir{
	Notice that as has been mentioned in the discussion following 
	equation \ref{GeneralBoltz},  
	such a sink term mimics the true physics deep inside the LC 
	(at depth larger than $\sim j_d$). 
	Thus, since a star that have entered those regions, typically does not scattered out of the LC, but rather remains there until its destruction.	     
	}
}
\end{itemize} 
Since this simulates the angular momentum evolution over time $\Tsim$,
it effectively gives the steady state profile up to angular
momentums
$\gtrsim\Jsim\equiv [\Tsim D(2j_{\LC})]^{1/2}$.

\subsection{Simulation parameters}\label{parameters}
The parameters of the simulation all in units of $j_{\LC}$ and $t_d/2$
are as follows:
\begin{itemize}
\item {$j_d$ - the typical scattering at the LCs surrounding, 
		which determents how empty the LC is.
		The value of $j_d$, set the rates of all scattering via its relation to $R_0$  
		(equation \ref{jd_def}).}	
\item {$j_{\min}$ - the minimal possible scattering.}
\item {$j_{\max}$ - the maximal possible initial angular momentum. 
	We use $j_{\max}$ also as the maximal possible scattering.  
	Ideally $j_{\max}\rightarrow \infty$ }
\item {$\Tsim\text{ or } \Jsim$ - 
	$\Tsim$ is the virtual time we let the simulation evolve. 
	We define $\Jsim\equiv  [\Tsim D(2j_{\LC})]^{1/2}$ and demand $j_{\LC}\ll \Jsim\ll j_{\max}$.}
\item {$N_p$ - Number of particles. This number should be as large as possible 
	in order to ensure large number of particles deep in the LC. }
\end{itemize}
We have preformed three simulations for the following parameters: 

\medskip
\begin{tabular}{c || c | c | c }
   & simulation A & simulation B & simulation C \\
\hline
\hline
  $j_{\min}$ & $10^{-2}$ & $10^{-2}$ & $10^{-1}$ \\
  $j_d$      & $8.8\cdot 10^{-2}$ & $8.8\cdot 10^{-2}$ & $2.3\cdot 10^{-1}$ \\
  $j_{\max} $& $10^2$ & $10^3$ & $10^2$  \\
  $\Jsim $& $12.4$ & $124$ & $14.6$ \\
  $N_{p} $ & $2\cdot 10^{10}$ & $6\cdot 10^{9}$ & $2\cdot 10^{10}$  \\
\end{tabular}

\medskip
Simulations A and B represent the same physics (same $j_{\min}, j_d$).
The main difference between those two, is at $\Tsim$ and correspondingly $\Jsim$ and at $N_p$. 
Simulation A is designed to give the details of the profile at the loss cone surrounding while simulation B gives an insight
on relatively large distances (the deviations from $f_{\fnull}$ and $f_{\fexp}$ in high $j$'s).
Simulation C demonstrates the case of a mildly empty loss cone. 

\subsection{results}
 
\begin{figure}[h]
    \centering
   \includegraphics[width=0.5\textwidth]{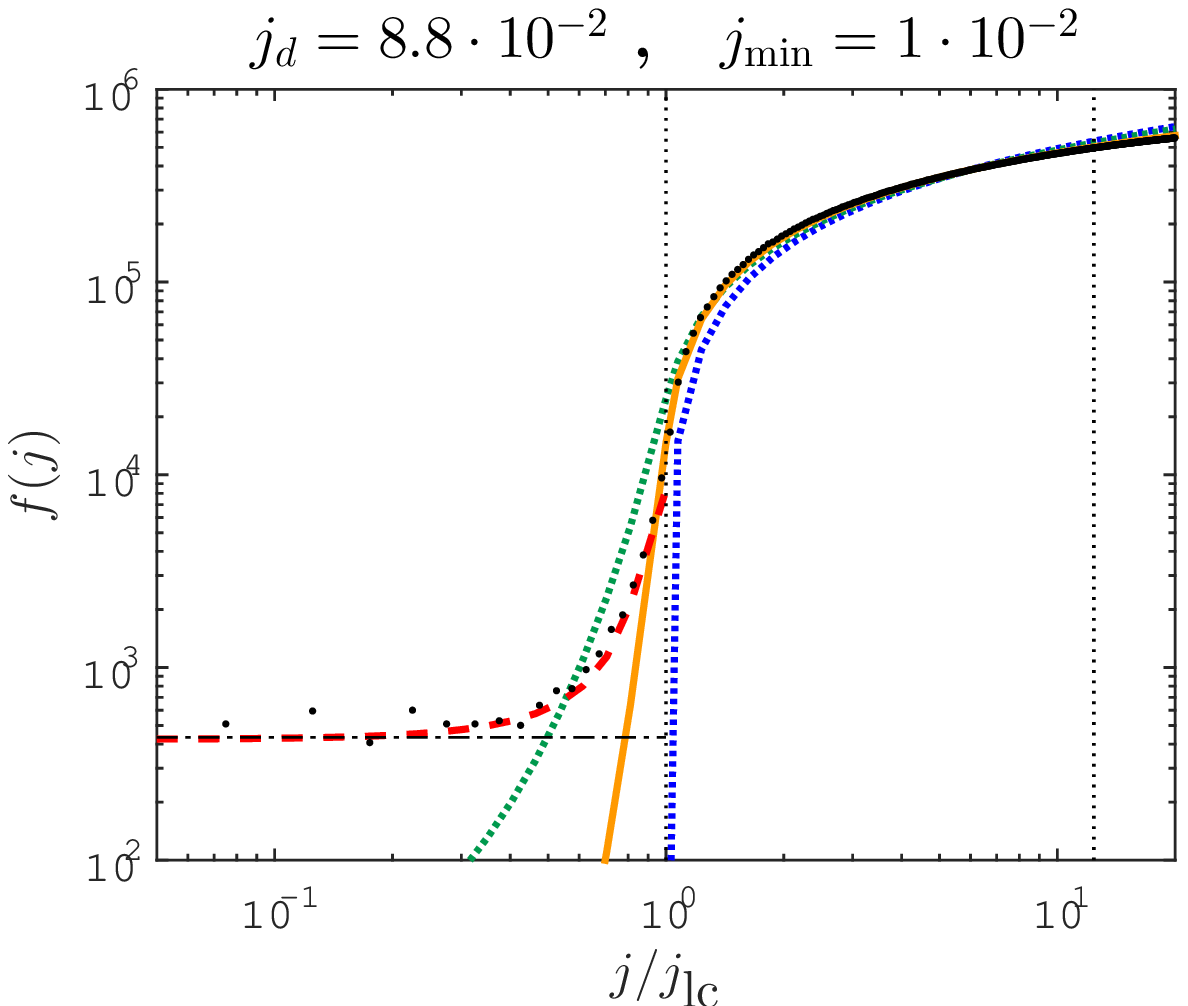}
	\caption{The numerical results (Black dots) of simulation A, 
		compared to the different analytical models:
		$f_{\fnull}$ in doted blue line,
		$f_{\fexp}$ in doted green line,
		$f_{\fdiff}$ in continues orange line
		and
		$f_{\os}$ in dashed red line.
		$f_{\os}^{deep}$, 
		\amir{the analytical approximation of $f_{\os}$ preesented in equation \ref{f1stepdeep}}
		is also presented (black dashed-doted line).
		As can be seen, the suggested model,
		which represented by the $f_{\fdiff}(j)$ outside the loss cone and close to its edge 
		and $f_{\os}(j)$ at depths $\gtrsim j_d$ into it (equations \ref{fdibar} and \ref{f1step}), 
		agrees with the data up to very good precision.
}
    \label{figsimA}
\end{figure}

\begin{figure}[h]
    \centering
    \includegraphics[width=0.5\textwidth]{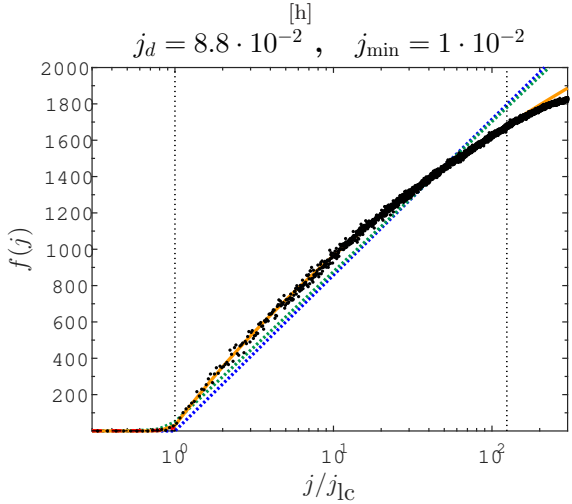}
	\caption{The numerical results of simulation B, compared to the different analytical models.
		The different models are marked with the same colors as in figure \ref{figsimA}.
		The relatively large $\Jsim$ ($\cong 10^2j_{\LC}$) provide clear disagreement between the constant diffusion coefficient model ($f_{\fnull}(j)$ or $f_{\fexp}$) 
		and the non-constant diffusion coefficient model ($f_{\fdiff}(j)$).  
		As expected, the simulation's result agrees with $f_{\fdiff}$. 
}
    \label{figsimB}
\end{figure}

\begin{figure}[h]
    \centering
    \includegraphics[width=0.5\textwidth]{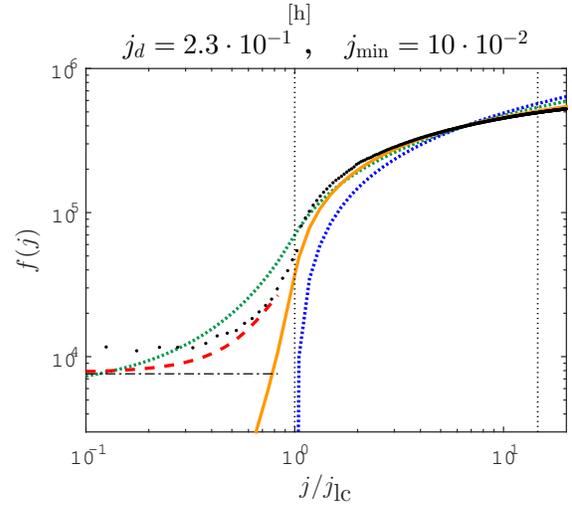}
	\caption{The numerical results of simulation C, compared to the different analytical models.
		The different models are marked with the same colors as in figure \ref{figsimA}.
		Since $j_d/j_{\LC}\cong 0.2$, the loss cone is only marginally empty. 
		Yet, there is better agreement between the result and the predicted model ($f_{\os}$ inside $f_{\fdiff}$ outside), than with any other option.  
		\amir{Notice that $f_{\os}^{deep}$, 
		the analytical approximation of $f_{\os}$ at $j\ll j_{\LC}$, describe $f_{\os}(j=0)$ well.}
}
    \label{figsimC}
\end{figure}

The results of the three simulations,
and the theoretical models for the relevant parameters, 
are all presented in figures 
\ref{figsimA}, \ref{figsimB} and \ref{figsimC}. 
In general, the suggested model,
which is represented by the $f_{\fdiff}(j)$ outside the loss cone and close to its edge 
and $f_{\os}(j)$ at depths $\gtrsim j_d$ into it (equations \ref{fdibar} and \ref{f1step}), 
agrees with the data.

Figure \ref{figsimA}, shows the occupancy distribution in the LC's vicinity.	
		Our main result is that inside the LC, at $j_{\LC}-j\gg j_d$, all the diffusive profiles 
		($f_{\fnull},f_{\fexp},f_{\fdiff}$) fail to describe the actual profile of 
		simulation A.
		As has been claimed in \S\ref{chapf1step} the profile in this region is 
		dominated by objects who have entered the LC via one large scattering
		and indeed $f_{\os}(j)$ describes the profile in those regions well.

As expected, the profile in the LC's edges surroundings is best described by $f_{\fdiff}(j)$.
$f_{\fnull}(j)$ 
under estimates the
profile at distance of $\sim j_d$ from the LC's edge,  
while $f_{\fexp}(j)$ overestimate
it. 
	Figure \ref{figsimA} also shows 
    a flattening of the profile in the high $j's$ ($j\gg j_{\LC}$) with respect to $f_{\fnull}$ or $f_{\fexp}$. 
	To better capture this feature,
	we have preformed another simulation (Simulation B)
	which has the same physical parameters ($j_d,j_{\min}$)
	but a much larger $\Jsim$ 
	(on the price of a much smaller number of particles).
	Figure \ref{figsimB} presents the results of this simulation. 
	Notice the profile deviate significantly from any constant diffusion coefficient model 
	($f_{\fnull}$ or $f_{\fexp}$). 
	$j$'s dependent diffusion coefficient on the other hand, 
	led us to $f_{\fdiff}(j)$ and that profile seems to agrees well with the actual profile. 
	
		Figure \ref{figsimC} present the results of simulation C. 
		It shows the applicability of our approximations in a more 
		marginal empty LC's case, were $j_d/j_{\LC}=0.23$.   
	
\section{Realistic examples}\label{realgalaxy} 

In sections  \ref{chapfdibar} and \ref{chapf1step} we have described two effects.
The first, is the flux reduction as a consequence of the lower diffusion coefficient (equation \ref{rate_factor})
and the second is the presence of objects deep inside the LC,
even if $j_d\ll j_{\LC}$.
The significance of the described effects, 
depends upon the values of $j_{\min},j_d$ and $j_{\LC}$,
we therefore discuss the
values of these parameters in realistic systems. 

\subsection{Parameters in an equal mass realistic systems}
Consider a massive black hole of mass $M_{\BH}$, 
surrounded by spherical bulge.
We assume that the massive black hole's sphere of influence contains 
$\sim M_{\BH}/m_0$ objects of mass $m_0$.

While experiencing a scattering of impact parameter $b$ with an object of mass $m_0$,
the typical object changes its angular momentum by $\delta j\sim j_{\circular}\cdot Gm_0/bv^2$.
Since scattering at the sphere of influence, $r\sim r_{\BH}$ dominate the scatterings, 
we find that the minimal possible scattering 
is:
\begin{equation}\label{jmin_real}
j_{\min}\sim \left(\frac{m_0}{M_{\BH}}\right)j_{\circular}\  \ .
\end{equation}   
The number of scatterings of size $j_{\min}$ during one dynamical time is roughly the number of the stars ($M_{\BH}/m_0$).
Therefore, the contribution from scatterings of size about $j_{\min}$ to the diffusion coefficient is 
$\sim (M_{\BH}/m_0) j_{\min}^2t_d^{-1}$. 
Since other step sizes have equal contribution we find from equation (\ref{jd_def})
\begin{equation}\label{jd_real}
j_d 
\sim 
\Lambda_{\LC}^{1/2} 
\left(\frac{m_0}{M_{\BH}}\right)^{1/2} j_{\circular}.
\end{equation}
The loss cone size, $j_{\LC}$, depends upon the ratio between the tidal radius, $r_t$,
and the radius of the sphere of influence, $r_{\BH}$: 
\begin{equation}
j_{\LC}\simeq\left(\frac{2r_t}{r_{\BH}}\right)^{1/2}j_{\circular}\ .
\end{equation}
where for tidal disruption event (TDE) of star of mass $m_*$ and radius $R_*$ 
\begin{equation}
r_t\simeq \left(\frac{M_\BH}{m_*}\right)^{1/3}R_*\ .
\end{equation}
Comparing $j_{\LC}$ with $j_d$
taking numerical factor exact for isothermal sphere
	\footnote{
		We have used $j_d=(\Lambda_{\LC}/\Lambda_{\max})^{1/2}j_2$
		where $j_2=(t_d/t_{\text{relax}})^{1/2}j_{\circular}$ and $t_{\text{relax}}$ 
		is the Chandrasekhar relaxation time $t_{\text{relax}}=0.34\sigma^3/G^2m_0\rho$ \citep{LS1977}.   
		Than, assuming an isothermal sphere, we have substituted $r_{\BH}=GM_{\BH}/\sigma^2$ and $\rho=M_\BH/2\pi r_{\BH}^3$
		\citep{M2003}.
	},
we find: 
\begin{equation}
\frac{j_d}{j_{\LC}}
\simeq
\frac{\Lambda_{\LC}}{2\cdot 0.34}
\frac{Gm_0m_*^{1/3}}{M_{\BH}^{1/3}R_*\sigma^2}
\quad.
\end{equation}

\amiraddF{Loss cone in this case turn out to be only marginally empty considering a typical main sequence star. 
For a main sequence star}
$R_*\simeq(m_*/m_\odot)^{0.8}R_{\odot}$
and an $M-\sigma$ relation of the form
$M_{BH} = 
M_0\left(\sigma/200\text{km s}^{-1}\right)^{\beta} $
where $M_0=1.32\cdot 10^8m_{\odot},\ \beta=4.24$ \citep{Msigma2009}
we find
\begin{equation}
\frac{j_d}{j_{\LC}}
\simeq
0.41\cdot
\left(\frac{\Lambda_{\LC}}{10}\right)^{1/2}
\left(\frac{m_*}{m_\odot}\right)^{-0.233}
\left(\frac{M_\BH}{10^8m_\odot}\right)^{-0.4}
\quad.
\end{equation}
In principle this may lead to empty LC at large $M_\BH$,
however, since the maximal $M_{\BH}$ for which a TDE occurs outside the horizon is
$\simeq 10^8m_{\odot}\cdot (m/m_\odot)^{0.7}$,
empty LC's scenario is 
\amiradd{only marginally}
relevant only for relatively massive stars and MBHs. 	
\amiraddF{
	However for Giants, empty LC scenario become relevant. 
	Since the radius of a giant of mass $m_\odot$,
	is $\gtrsim  100R_\odot$, 
	$r_t$ also grows by factor of $\sim 100$ and $j_{\LC}$ grows by factor of $10$. 
	This makes $j_d/j_{\LC}\sim 0.04$ (LC emptier than presented in figure \ref{figsimA})}   
	\footnote{
		\amiraddF{
			Notice however that although $j_d/j_{\LC}$ in the simulation is approximately similar to that of the realistic system,
			$j_{\min}/j_{\circular}$ in the simulation is much larger than in the reality (reduction of $j_{\min}$ and enhancement of the maximal $j$ for which the system is relaxed is numerically expansive).
			As a result, $\Lambmax$ and $\Lambzero$ in a real system are larger by a factor of $\simeq 2$ than in the simulation. 
		}}

\amiradd{
For binary's breakup which leads to hypervelocity stars (HVSs) production, 
$r_t\simeq (M_{\BH}/2m_*)^{1/3}a$ where $a$ is the binary's separation and $2m_*$ is the binary's mass.
In this case: 
\begin{equation}
\frac{j_d}{j_{\LC}}
\simeq
0.25\cdot
\left(\frac{\Lambda_{\LC}}{10}\right)^{1/2}
\left(\frac{m_*}{m_\odot}\right)^{-0.233}
\left(\frac{M_\BH}{10^8m_\odot}\right)^{-0.4}
\left(\frac{a}{2R_*}\right)^{-1/2}	
\end{equation}	
and therefore the LC may be empty also for low mass MBHs such as Sagittarius A* or for low mass stars.
For instance, 
a binary of B stars ($4m_\odot$ each, see \cite{BGK2009}) and separation of $15R_*$, 
which interact with an MBH of mass $4.3\cdot 10^6m_{\odot}$, 
has a mildly empty LC with $j_d/j_{\LC}\simeq 0.23$ (same as in \ref{figsimC});
and a binary of two solar type stars with separation of $15R_\odot$, 
interacting with Andromeda's MBH ($M_\BH\simeq 1.2\cdot 10^8m_\odot$) has 
$j_d/j_{\LC}\simeq 8.4\cdot 10^{-2}$ (smaller than $j_d/j_{\LC}$ of figure \ref{figsimA}). 
}

\subsection{LC filling rate reduction in a realistic system}\label{Section_reduction}
Notice that the relation between $j_{\min}$ and $j_d$ as given by 
equations \ref{jmin_real} and \ref{jd_real},
restrict the LC's influx rate reduction presented in equation \ref{rate_factor}.
According to equation \ref{rate_factor},
the flux reduction is $\sim \Lambzero/\Lambda_{\circular}$
in the regime we later find to be relevant of $\Lambzero\gtrsim \Lambmax$.
Since our model only assumes empty LC
we require $j_d \ll j_{\LC}$ 
and obtain:
%
%
\begin{subequations}
\begin{align}
&\Lambzero=
\frac{1}{2}\ln
\left[\Lambda_{\LC}\left(
\frac{M_{\BH}}{m_0}\right)\right]
+\ln\left(\frac{j_{\LC}}{j_d}\right)
\\
&\Lambda_{\circular}=
\ln\left(\frac{M_{\BH}}{m_0}\right)<2\Lambzero
\end{align}
\end{subequations}
therefore, in a system in which all objects have the same mass,
the influx reduction (for $\ln(j_{\LC}/j_{\min})\ll \ln(M_{\BH}/m_0)$) 
is about a factor of $2$.
\section{summery and applications}

We discussed the effect of large scatterings on the occupancy distribution of stars in the presence of an empty loss cone. 
We expect the occupancy distribution to be of the form of 
$f_{\fdiff}$ (equation \ref{fdibar}) outside the LC and at depth $\lesssim j_d$ into it, and 
$f_{\os}$ (equation \ref{f1step}) at larger depths inside the LC. 
Those expectations were deduced from an analytical model and verified using Monte Carlo simulations.   

Our solution, contain two major differences compared to the standard Fokker-Planck approach \citep{LS1977,CK1978,MT1999,MerrittBook2013}. 

First, at large angular momentums $j\gtrsim j_{\LC}^2/j_{\min}$ (if those exist) the profile is not $\propto\ln(j/j_{\LC})$ as expected from a Fokker-Planck approximation, but 
\amiradd{flatter, with an asymptotic behavior $\propto \ln[\ln(j/j_{\LC})]$ (equation \ref{fdibar}).} 
This flatter profile make the flux into low angular momentums a little lower.
More accurately, if $j_{\circular}<j_{\LC}^2/j_{\min}$ the Coulomb logarithm 
$\ln(j_{\max}/j_{\min})$ should be replaced with $\ln(j_{\LC}/j_{\min})$ while calculating the flux,
and if $j_\circular>j_{\LC}^2/j_{\min}$ the rate is even slower than that (equation \ref{rate_factor}). 
This effect reduces the theoretical 
loss cone filling rate of the Milky way by a factor of $\approx 2$ for the case of HVS
where only two-body relaxation in spherical symmetry is take into account.
   
Second, empty loss cone is much less empty at depth $\gg j_d$ than expected from any kind of diffusive model.
According to diffusive approximations,
the profile inside the loss cone decays exponentially over a scale of 
$\sim j_d$ (similar to $j_2$ of \cite{CK1978}).
See our equations (\ref{fdelay}) or (\ref{fdibar}) or the classical work of \cite{CK1978}.
However, according to our new understanding, large scattering uniformly populate the loss cone. The profile deep inside the loss cone does not decay exponentially but is rather constant at a value of $\sim f(j_{\circular})(j_d/j_{\LC})^2\ln^{-2}(j_{\LC/j_{\min}})$.  

This second effect may have some observational significance. 
Let us discuss the nature of tidal events
\amiraddF{of giant stars 
in which the LC for tidal disruption event is empty 
}.
The hardness of tidal disruption events are characterized by the parameter:	 
		\begin{equation}
		\beta\equiv\frac{\text{tidal radius}}{\text{peri-astron distance}}=
		\left(\frac{j_{\LC}}{j}\right)^2\ \ \ .
		\end{equation}
		\medskip	
	Previous analysis based on the diffusion approximation  
	predicts the typical tidal event to have $\beta-1 \sim j_d/j_{\LC}$,
	were large $\beta$'s tidal events are exponentially rare with probability $\sim \exp\{-(\beta-1)j_{\LC}/j_d\}$.
	Instead, we find that large scatterings populate the distribution deep in the LC, making it lower only by factor of  $(j_d/j_{\LC})\ln^{-1}(j_{\LC}/j_{\min})$ compared to the occupancy distribution on the LC's edge. 
	Since the typical width of the edge is $\sim j_d$,  
	the total number of particles on the edge is only $\sim\ln(j_{\LC}/j_{\min})$ 
	higher than the number of particles at high $\beta$'s.
	As a result, the typical fraction of high $\beta$'s tidal events out of all tidal 
	events is $\sim\ln^{-1}(j_{\LC}/j_{\min})$.
	Moreover, since the profile deep inside the LC is roughly flat, 
		the probability of high $\beta$ events does not decays exponentially
		but scales as the surface within $j<j_{\LC}\beta^{-1/2}$, i.e as $\beta^{-1}$.
		
	This prediction may be verified observationally  
if deeply penetrating tidal events look different from
shallow ones. 

\section*{Acknowledgements}

This research was partially supported by iCore, ISF and ISA grant

\appendix

\section{Which scatterings are important ?}\label{appendix_large_scatterings}
In section \ref{chapfdibar}, 
we argue that scatterings $\gtrsim j-j_{\LC}$
are less dominant outside the LC, and may be neglected.
This assumption lead us to $j$'s dependency of the diffusion coefficient (equation \ref{Dfuncj}).  
In this section we explain this assumption 

We start with estimating the total contribution 
of scatterings of size $\jscat$
to the net influx toward angular momentums lower than $\jtarg$
in the case $\jscat\ll \jtarg-j_{\LC}$
as well as in the case where $\jscat\gg \jtarg-j_{\LC}$.
For simplicity, we assume $\jtarg\gg j_{\LC}$ ($\jtarg\sim \jtarg-j_{\LC}$). However, same principles hold also for the case of $\jtarg-j_{\LC}\lesssim j_{\LC}$.   

For small scatterings, $\jscat\ll \jtarg$,
the diffusion approximation holds
and therefore
\begin{equation}
\gamma_{net}^{\sma}(\jscat,\jtarg)=
2\pi \jtarg D_{\jscat}\left.\frac{\pd f}{\pd j}\right|_{\jtarg}\sim R_0f(\jtarg)
\end{equation} 
where $D_{\jtarg}$ is the contribution to the diffusion coefficient of scatterings of size $\sim \jtarg$ which satisfies 
$D_{\jtarg}\sim R_0$.

For large 
scatterings, $\jscat\gg \jtarg$,
the diffusive approximation does not holds and the net influx have to be estimated from comparison of the influx and the outflux.
The influx into $j<\jtarg$ by scatterings of size $\sim \jscat$, is approximately the product of
the rate of $\jscat$ scatterings ($\sim R_0\jscat^{-2}$),
the number of particle at $\jscat$ surroundings ($\jscat^2f(\jscat)$)
and the probability that scattering of size $\jscat$ 
scatters an object into a sling-shot of size $\jscat\ll \jtarg$ ($(\jtarg/\jscat)^2$).
Therefore the influx into $j<\jtarg$ by scatterings of size $\sim \jscat$ is
\begin{equation}
\gamma_{in}^{\lar}(\jscat,\jtarg)\sim 
R_0\left(\frac{\jtarg}{\jscat}\right)^2f(\jscat)\ \ .
\end{equation} 
Similar estimation of the outflow from angular momentums $<\jtarg$ 
(without the probability term which in this case is unity) gives:
\begin{equation}
\gamma_{out}^{\lar}(\jscat,\jtarg)\sim 
R_0\left(\frac{\jtarg}{\jscat}\right)^2f(\jtarg)\ \ .
\end{equation}
Since $\jtarg<\jscat$, $f(\jtarg)$ is necessarily $\leq f(\jscat)$.
Therefore $\gamma_{in}^{\lar}(\jscat,\jtarg)$ is an upper limit for the net influx:
\begin{equation}
\gamma_{net}^{\lar}(\jscat,\jtarg)\lesssim 
R_0\left(\frac{\jtarg}{\jscat}\right)^2f(\jscat)\ \ .
\end{equation}

In order to understand which scattering dominate the profile. 
we would like to compare for each $\jtarg$,
the total contribution of scatterings smaller than $\jtarg$ ($\Gamma_{net}^{\sma}(\jtarg)$),
with that of scatterings larger than $\jtarg$ ($\Gamma_{net}^{\lar}(\jscat)$):
\begin{subequations}
	\begin{align}
	&\begin{array}{l l}
	\Gamma_{net}^{\sma}(\jtarg)\equiv &
	\sum\limits_{\jscat\lesssim\jtarg}\gamma_{net}^{\sma}(\jtarg,\jscat)
	\sim\\
	& R_0f(\jtarg)\ln\left(\frac{\jtarg}{j_{\min}}\right)
	\end{array}
\\
	&\begin{array}{l l}
	\Gamma_{net}^{\lar}(\jtarg)\equiv &
	\sum\limits_{\jscat\gtrsim\jtarg}\gamma_{net}^{\lar}(\jtarg,\jscat)
	\sim
	\\
	& 
	R_0\jtarg^2\cdot\max\{\left.j^{-2}f(j)\right|_{j>\jtarg}\}
	\end{array}
	\end{align}
\end{subequations}

At this stage we are ready to formulate properly the question of 
"which scattering dominate the profile?".
There are two possibilities:
\begin{itemize}
	\item {If $f(j)$ grows slower than $\propto j^2$ in $\jtarg$'s neighborhood,
		than $\max\{\gamma_{net}^{\lar}\}=\gamma_{net}^{\lar}(\jtarg\sim\jscat)$
		In this case all scatterings at sizes $<\jtarg$ has equal contribution,
		while larger scatterings has smaller contribution. 
		}
	\item{ On the other hand, 
		if $f(j)$ grows faster than $\propto j^2$ in $\jtarg$'s neighborhood,
		than 
		$\max\{\gamma_{net}^{\lar}\}>
		\gamma_{net}^{\lar}(\jtarg\sim\jscat)
		\sim \Gamma_{net}^{\sma}(\jtarg)$.
		In this case there is some scale$>\jtarg$, 
		which dominate the scatterings everywhere. 
		}	
\end{itemize}
Understanding which scattering dominate the profile is therefore not simple, since it requires some assumptions about $f(j)$.
We therefore discuss this question
under the assumption that small scatterings dominate the profile, 
and afterwards 
under the assumption that large scatterings are dominant.
We show that the assumption of small scattering dominance is self consistent and give arise to a steady state profile, 
while assumption of large scattering dominance, is not.

If small scattering ($\jscat\ll \jtarg$) dominate the profile,
it is set by diffusion with diffusion coefficient of the form \ref{Dfuncj}.
In this case the discussion of \S\ref{chapfdibar} is relevant
and $f(j)$ has a steady state form of $f_\fdiff(j)$ (equation \ref{fdibar}).
Since $f_\fdiff(j)$ grows slower than $\propto j^2$,
small scatterings are dominant and the argument is self-consistent.

On the other hand,
if large scattering dominate the profile, the profile has to grow faster than 
$\propto j^2$ at $\jtarg$'s neighborhood. 
Lets assume for instance that it grows as $f(j)=f(\jtarg)(j/\jtarg)^N$ ($N>2$) up to some angular momentum $j_{\gamma}\gg \jtarg$, for which $j^{-2}f(j)$ is maximal.
Under this assumption, the influx rate into $j<\jtarg$ is:
$R_0f(\jtarg)(j_{\gamma}/\jtarg)^{N-2}$.
This dependency of the influx rate upon $\jtarg$ 
prevent zero divergence and as a result, the profile can not be in steady state.

Steady-state profile therefore grantee dominance of the small scatterings outside the LC: 
In some particular $\jtarg$, all scatterings of size 
$\lesssim \jtarg $ have the same contribution to the diffusion coefficient while scatterings $\gtrsim\jtarg $ can be ignored.

\section{Accelerating the numerical scheme}\label{improve}
As a working example, 
we consider a system with $j_d\sim 10^{-1}$, $j_{\min}\sim 10^{-2}$ (while $j_{\LC}\equiv 1$), evolved enough so that $\Jsim\sim 10$. 
Since we are interested in the occupancy distribution inside the loss cone, we require that  the total number of particles in its inner half (with surface of $\pi 0.5^2 \sim 1$) 
would be at list $10$ particles for sufficient statistical accuracy. This means a particle density of $\sim 10$ deep in the LC.

The largest angular momentum in the simulation, $j_{\max}$ has to be a few times larger than $\Jsim$ and we take $j_{\max}=30$.
The particle density at $j\sim j_{\max}$ is higher by factor of   
$\sim (j_{\LC}/j_d)^2\ln^{2}(j_{\LC}/j_{\min})\sim 2.5\cdot 10^3$ compared to the density deep in the LC
which was assumed to be $\sim 10 $ and therefore,
the total number of particles is $10^8$.

Relaxation has to take place over a scale of $\Jsim\sim 10$ meaning that each particle experience $\sim (\Jsim/j_{\min})^2\ln^{-1}(\Jsim/j_{\min})\sim 10^5$ steps
which, leads to a total number of $\sim 10^{13}$ computations.
We therefore introduce two major improvements to accelerate the numerical scheme. 

\subsection{Focusing our efforts on the loss cone surrounding}
Most of the scatterings are small and occurs far from the LC.
In these regions, calculating the frequent scattering of size $j_{\min}$ is not
necessary. We therefore replace those frequent scatterings with a few larger ones,
which are still smaller than the typical angular momentum over which $f$ is changing.
This is done as follows. We set the minimal size of the scatterings at angular momentum $j$ to be 
\begin{equation}
j_s(j)= j_{\min} \cdot \max \left\{1,\frac{j}{2j_{\LC}}\right\}
\end{equation}
and replaced $R(\delj)$, the rate of each scatterings to:
\begin{equation}\label{Rjeff}
R_{\text{eff}}(\delj) = \xi(\delj)R(\delj) 
\end{equation}
with
\begin{equation}
\xi(\delj) = \left\{
\begin{array}{c c }
1+\ln\left(\frac{j_s}{j_{\min}}\right) & \text{for $j_s<\delj<ej_s$} \\
\\
1 & \text{for $ej_s<\delj$}
\end{array}
\right.
\end{equation} 
Those definitions reduced significantly the number of calculations on one hand, 
and on the other hand mimic well the real behavior:  
the total contribution of the small scatterings to the diffusion does not changed
as well as the rate of large scatterings. 
Moreover, the scatterings' rate close to the LC has not been changed at all. 
In the working example above, this reduces the number of computations by three
orders of magnitude.

In order to implement $R_{\text{eff}}(\delj)$ in the simulation, 
The time length attributed to each step has to be recalculated 
(it is no longer $t_{\min}$ since steps of size larger than $j_s$ are more rare than once every $t_{\min}$). 
We therefore attribute to each time step a period of $t_s(j_s)$ where $t_s$ is set so that the rate of scatterings of size $>ej_s$ would not be influenced by the presented adaptation and  we find $t_s$ as a function of $j_s$ to be: 
\begin{equation}
\begin{array}{l l }
t_{s}
= &
\frac
{\int_{j_{\min}}^{j_{\max}} d\delj \ \delj^{-3}}
{\int_{j_{s}}^{j_{\max}} d\delj \ \xi(\delj) \delj^{-3}}
t_{\min}
=
\\ &
\left(\frac{j_s}{j_{\min}}\right)^{2}
\left[\frac
{1-\left(j_{\max}/j_{\min}\right)^{-2}}
{(1-e^{-2})\ln\left(\frac{j_s}{j_{\min}}\right)+1 -\left(j_{\max}/j_{\min}\right)^{-2}}
\right]t_{\min}
\end{array}
\end{equation}
Notice that for $j_s=j_{\min}$ we get $t_s=t_{\min}$ as expected. 

\subsection{Using relaxed profile to account for more particles}
After time $\Tsim$, the profile is already relaxed up to angular momentums $\sim \Jsim$. 
However, if after achieving this steady state we let the system evolve for a period of another $t_{\text{add}}\equiv(j_{\LC}/\Jsim)^2\Tsim$, all particles would change their location by typically angular momentum $\sim j_{\LC}$. 
This means that there is no correlation between the two situations on scale of $\lesssim j_{\LC}$ and we can use this result as an independent profile.

Repeating this process over and over again, we may increase our statistics on the same physical condition without creating dependence between the different repeats.
This may go on and on as long as $N_{\text{rep}}\cdot t_{\text{add}} \lesssim t_{sim} $ where $N_{\text{rep}}$ is the number of repeats. 

\bibliographystyle{mn2e}
\bibliography{howEmpty}
\end{document}